\documentclass[10pt,conference]{IEEEtran}
\IEEEoverridecommandlockouts
\usepackage{pifont}          
\usepackage{xcolor}          
\usepackage{colortbl}        
\usepackage{multirow}        
\usepackage{cite}
\usepackage{amsmath,amssymb,amsfonts}
\usepackage{algorithmic}
\usepackage{algorithm}    
\usepackage{url}
\usepackage{booktabs}
\usepackage{xurl}
\usepackage{caption}
\usepackage{subcaption}
\usepackage{multicol}
\usepackage{algorithmic}
\usepackage{graphicx}
\usepackage{textcomp}
\usepackage{xcolor}
\usepackage{color}
\usepackage{listings}
\usepackage{xspace}
\usepackage{booktabs}
\usepackage{array}
\usepackage{enumitem}
\usepackage{tcolorbox}
\usepackage{array}

\def\BibTeX{{\rm B\kern-.05em{\sc i\kern-.025em b}\kern-.08em
    T\kern-.1667em\lower.7ex\hbox{E}\kern-.125emX}}
\begin{document}

\newcommand{\CBrush}{\textcolor[RGB]{84,130,53}{\ding{51}}}
\newcommand{\XBrush}{\textcolor[RGB]{176,35,24}{\ding{55}}}
\newcommand{\PBrush}{\textcolor[RGB]{255,165,0}{\ding{73}}} 

\title{JSidentify-V2: Leveraging Dynamic Memory Fingerprinting for  Mini-Game Plagiarism Detection }

\author{
\IEEEauthorblockN{Zhihao Li\IEEEauthorrefmark{2}\IEEEauthorrefmark{4}, Chaozheng Wang\IEEEauthorrefmark{2}\IEEEauthorrefmark{3}\IEEEauthorrefmark{4}, Zongjie Li\IEEEauthorrefmark{5}\IEEEauthorrefmark{6}, Xinyong Peng\IEEEauthorrefmark{2}, Qun Xia\IEEEauthorrefmark{2}, Haochuan Lu\IEEEauthorrefmark{2}, Ting Xiong\IEEEauthorrefmark{2}, \\
Shuzheng Gao\IEEEauthorrefmark{3}, Cuiyun Gao\IEEEauthorrefmark{3}, Shuai Wang\IEEEauthorrefmark{6}, Yuetang Deng\IEEEauthorrefmark{2}, Huafeng Ma\IEEEauthorrefmark{2}
}
\IEEEauthorblockA{\IEEEauthorrefmark{2} \textit{Tencent Inc.} Shenzhen, China \\}
\IEEEauthorblockA{\IEEEauthorrefmark{3} \textit{The Chinese University of Hong Kong} Hong Kong, China \\}
\IEEEauthorblockA{\IEEEauthorrefmark{6} \textit{The Hong Kong University of Science and Technology} Hong Kong, China \\}
\IEEEauthorblockA{\IEEEauthorrefmark{4} Equal contribution \\}
\IEEEauthorblockA{\IEEEauthorrefmark{5} Corresponding author}
}

\maketitle

\begin{abstract}


The explosive growth of mini-game platforms has led to widespread code plagiarism, where malicious users access popular games' source code and republish them with modifications. While existing static analysis tools can detect simple obfuscation techniques like variable renaming and dead code injection, they fail against sophisticated deep obfuscation methods such as encrypted code with local or cloud-based decryption keys that completely destroy code structure and render traditional Abstract Syntax Tree analysis ineffective. To address these challenges, we present JSidentify-V2, a novel dynamic analysis framework that detects mini-game plagiarism by capturing memory invariants during program execution. Our key insight is that while obfuscation can severely distort static code characteristics, runtime memory behavior patterns remain relatively stable. JSidentify-V2 employs a four-stage pipeline: (1) static pre-analysis and instrumentation to identify potential memory invariants, (2) adaptive hot object slicing to maximize execution coverage of critical code segments, (3) Memory Dependency Graph construction to represent behavioral fingerprints resilient to obfuscation, and (4) graph-based similarity analysis for plagiarism detection.

We evaluate JSidentify-V2 against eight obfuscation methods on a comprehensive dataset of 1,200 mini-games. Our framework achieves over 90\% similarity detection across all tested obfuscation techniques, maintaining high accuracy even against advanced decryption-based methods where existing tools achieve near 0\% detection rates. In real-world deployment, JSidentify-V2 achieves 100\% precision and 99.8\% recall while delivering an 8$\times$ speedup compared to previous methods. Our production deployment demonstrates that plagiarism complaints have decreased by over 80\%, proving JSidentify-V2's effectiveness in protecting intellectual property rights in mini-game ecosystems.

\end{abstract}

\begin{IEEEkeywords}
Mini-games, JavaScript, Plagiarism detection
\end{IEEEkeywords}

\section{Introduction}
The proliferation of smartphone technology has fundamentally transformed the mobile gaming landscape, driving unprecedented growth in this sector over the past decade. According to industry analytics, global mobile game revenue is projected to reach \$135 billion across app stores \cite{data}. This immense profitability has not only fostered a burgeoning ecosystem of game development but has also led to the proliferation of various mini-game platforms. These platforms distinguish themselves from traditional mobile game applications by embedding lightweight games directly within existing mobile applications, thereby eliminating the need for separate downloads or installations. This ``plug-and-play'' functionality substantially enhances user engagement and has attracted a vast user base, alongside a rapidly expanding community of developers and platform providers. A prime example of such a successful ecosystem is the Tencent \textit{WeChat} mini-game platform, which has garnered hundreds of thousands of developers and hosts hundreds of thousands of mini-games, serving a user base exceeding one billion and boasting 500 million monthly active users \cite{wang2023unified}.

\begin{figure*}[t]
  \centering
  \includegraphics[width=0.75\linewidth]{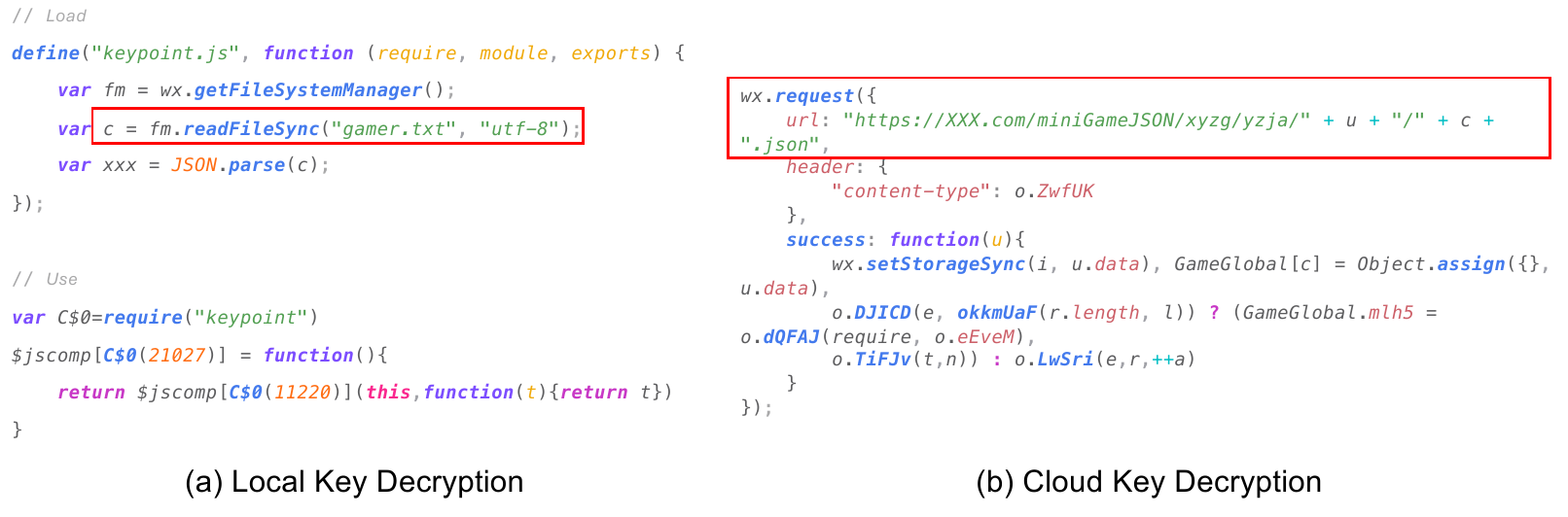}
  \caption{Examples of advanced deep obfuscation techniques that static analysis fails to detect. In (a) Local Key Decryption, the core code is encrypted, and the decryption key is retrieved from a local file. In (b) Cloud Key Decryption, the key is fetched from a remote server via a network request, with decryption happening dynamically.}
  \label{fig:example}
\end{figure*}


While the platform provided by \textit{WeChat} has undeniably been a pivotal force in accelerating the development and adoption of mini-games, it has concurrently introduced significant challenges, particularly concerning intellectual property infringement. 
The inherent characteristics of mini-games, often developed with simplified architectures and readily accessible codebases (e.g., JavaScript), render them particularly vulnerable to various forms of plagiarism. Malicious users frequently exploit the success of popular mini-games by illicitly acquiring their source code and repackaging it with modifications ranging from superficial changes such as asset replacements to sophisticated obfuscation techniques designed to evade detection. These plagiarized versions are then uploaded to the platform as competing products, directly infringing upon the intellectual property rights of original creators while diverting potential revenue streams away from legitimate developers. Such plagiarism not only undermines the creative efforts and economic interests of original developers but also poses broader threats to the platform's ecosystem integrity and user trust.

To address the escalating issue of code plagiarism in mini-games, the \textit{WeChat} development team previously proposed JSidentify~\cite{xia2020jsidentify}, a framework designed for the detection of plagiarized mini-game code. JSidentify primarily leverages static analysis techniques to identify common obfuscation methods such as code restructuring applied to JavaScript code. However, recent developments over the past two to three years have revealed the emergence of significantly more challenging obfuscation techniques that effectively bypass JSidentify's detection capabilities. Figure~\ref{fig:example} illustrates two typical examples of these advanced obfuscation strategies: (a) local key decryption, where plagiarists encrypt the core code and employ file-based key retrieval, and (b) cloud key decryption, where decryption occurs dynamically via network communication. These deep obfuscation techniques severely corrupt the structural integrity of the code, profoundly damaging Abstract Syntax Trees (ASTs) and character-based features. Consequently, static analysis-based de-obfuscation approaches fail due to the critical loss of information, rendering them ineffective against these sophisticated attacks.

In light of the limitations faced by existing static analysis methods when confronted with these sophisticated deep obfuscation techniques, this work explores the necessity and feasibility of incorporating dynamic analysis. The fundamental insight driving this approach is that while obfuscation can severely disrupt the static structural characteristics of code, the program's runtime memory behavior patterns tend to remain relatively stable. Building upon this observation, this paper proposes a novel detection framework that synergistically combines dynamic and static analysis, designed to identify plagiarism by capturing and analyzing the program's memory state during execution. The core concept of this methodology lies in identifying \textit{memory invariants}, which are runtime artifacts whose values or relational properties remain consistent across different executions of programs sharing the same core logic, even under deep obfuscation. These memory invariants are then organized into \textit{Memory Dependency Graphs (MDGs)} that serve as robust behavioral fingerprints, enabling effective plagiarism detection. 


Building upon the aforementioned arguments for the viability of dynamic analysis, we introduce JSidentify-V2, a new generation anti-plagiarism framework for mini-games. Our proposed framework employs a novel four-stage pipeline approach, offering a systematic solution to the complex challenge of deep obfuscation. The framework integrates four core components: (1) \textbf{Static Pre-analysis and Instrumentation} that identifies potential memory invariants and inserts monitoring probes; (2) \textbf{Adaptive Hot Object Slicing} that intelligently enhances execution coverage by focusing on critical program entities and domain-specific patterns; (3) \textbf{Memory Dependency Graph (MDG) Construction} that organizes collected invariant data into structured behavioral fingerprints; and (4) \textbf{Graph-based Similarity Analysis} that compares MDGs to detect plagiarism. The key innovation lies in our adaptive hot object slicing strategy, which systematically targets frequently executed functions and semantically important identifiers to maximize coverage of obfuscated code segments. This approach effectively captures behavioral invariants that remain stable across sophisticated obfuscation transformations, providing a robust and resilient solution to plagiarism detection.

We comprehensively evaluate JSidentify-V2's effectiveness against various obfuscation methods, demonstrating superior performance over existing approaches with over 90\% similarity detection across all eight tested obfuscation techniques. Remarkably, for advanced methods such as Local Key Decryption and Cloud Key Decryption, where the previous best baseline JSidentify achieves only around 5\% detection rate, our approach maintains consistently high performance. Furthermore, on our constructed dataset of 1,200 mini-games comprising 500 plagiarized pairs and 200 unrelated games, JSidentify-V2 achieves 100\% precision and 99.8\% recall, substantially outperforming all existing methods. Finally, by flexibly combining static and dynamic analysis, our approach delivers not only high accuracy but also remarkable efficiency, requiring only one-eighth the detection time of JSidentify while maintaining superior detection capabilities.

We summarize our contribution as follows:

\begin{itemize}
\item We report advanced obfuscation techniques that have emerged in the past two years, including local key and cloud key-based decryption methods, and demonstrate that these techniques can completely bypass all existing plagiarism detection tools.

\item We evolve a novel dynamic plagiarism detection framework, JSidentify-V2, which combines static analysis with dynamic execution to construct memory invariant values and build memory dependency graphs, enabling code fingerprint analysis even when the AST structure is completely destroyed.

\item Experimental results demonstrate that our method outperforms all baselines across eight different obfuscation techniques, achieving 99.8\% recall and 100\% precision while maintaining fast detection speed.
\end{itemize}

\section{Background}

\subsection{Obfuscation Methods}\label{sec:obfuscation}

Obfuscation techniques are commonly employed to hide code logic and evade plagiarism detection~\cite{collberg1997taxonomy,roy2007survey,ceccato2009effectiveness,schrittwieser2016protecting}. We evaluate our approach against eight obfuscation methods that represent different levels of complexity and detection difficulty.

\textbf{Identifier Modifications (IM)} replaces meaningful variable and function names with meaningless identifiers, such as converting \texttt{calculateSum} to \texttt{a} or random strings like \texttt{\_0x1a2b}. This technique aims to remove semantic information while preserving code functionality.

\textbf{Dead Code Injection (DCI)} inserts non-functional code segments that do not affect program execution but increase code complexity. These dummy statements and unreachable code blocks are designed to confuse static analysis tools and obscure the actual program logic.

\textbf{Control Flow Flattening (CFF)} restructures the program's control flow by converting nested control structures into flat switch-case statements or dispatcher patterns. This transformation makes it difficult to understand the original program flow and logic sequence.

\textbf{Nested Function (NF)} wraps code segments within multiple layers of function calls and closures, creating deep nesting structures that complicate code analysis. This technique often combines with scope manipulation to further obscure variable relationships.

\textbf{String Splitting (SS)} divides string literals into fragments that are concatenated at runtime, such as transforming \texttt{"Hello World"} into \texttt{"Hel" + "lo " + "Wor" + "ld"}. This method hides string constants from simple pattern matching.

\textbf{String Array Encoding (SAE)} converts string literals into encoded arrays with index-based access, often using Base64 or custom encoding schemes. For example, strings are stored in an encoded array and accessed through decoding functions during execution.

Recently, two advanced obfuscation methods have emerged that pose significant challenges to static analysis approaches. \textbf{Local Key Decryption (LKD)} encrypts code segments and embeds decryption keys within local files or configuration data, requiring runtime key extraction for code decryption. \textbf{Cloud Key Decryption (CKD)} takes this further by storing decryption keys on remote servers, making the code completely unanalyzable without network access and server authorization. These dynamic obfuscation techniques render traditional static analysis methods ineffective, as the actual code logic remains encrypted until runtime execution.

\subsection{Related Work}
Software plagiarism detection aims to identify unauthorized code copying~\cite{roy2007survey,li2022unleashing,rattan2013software,chae2013software,zhang2014program}. Effective plagiarism detection should account for different-level obfuscation, from basic Type I Identical Clones to Type IV semantically equivalent code with different syntactic structures \cite{xia2020jsidentify}. Various approaches have been developed including textual and lexical methods, AST-based methods~\cite{baxter1998clone}, and Program Dependence Graph (PDG)-based approaches~\cite{krinke2001identifying,gabel2008scalable}. Tools such as Simian~\cite{Simian} and MOSS~\cite{schleimer2003winnowing} are effective for detecting identical clones but struggle with gapped or obfuscated clones. AST-based tools like Jsinpsect~\cite{jsinspect} perform well for Type I and II clones but face challenges with more complex clone types. PDG-based approaches like JSCD (safe)~\cite{jhi2011value}, while efficient, often suffer from low precision and scalability issues. The previous framework JSidentify~\cite{xia2020jsidentify} leverages static analysis techniques to identify common obfuscation methods and largely outperform these methods. But it cannot deal with recent advanced obfuscation strategies like encryption. These challenges highlight the need for more robust detection methods that can handle advanced obfuscated code and various clone types effectively.

Moreover, large language models (LLMs) have been successfully applied to a wide array of code-related tasks, including code generation~\cite{gu2023llm,li2025api,wang2023codet52,shrivastava2023repository}, code completion~\cite{svyatkovskiy2019pythia,li2022cctest,bavarian2022efficient}, vulnerability detection~\cite{le2024study,wang2023reef,akuthota2023vulnerability}, and fuzzing~\cite{zhang2025low,jiang2024fuzzing,xia2024fuzz4all}. Some research even focuses on protecting the intellectual property of LLM-generated code through techniques like watermarking~\cite{li2023protecting,lee2023wrote}. Despite their power, LLMs have not been adopted for plagiarism detection in the mini-game ecosystem due to several practical limitations. As mini-games become increasingly complex with sophisticated features, their codebase size has grown substantially, often easily exceeding the million-token context limits of current LLMs~\cite{jaech2024openai,guo2025deepseek,touvron2023llama}. Additionally, the prohibitive computational costs render LLM-based analysis economically unfeasible for large-scale plagiarism detection in the mini-game ecosystem, which involves millions of daily comparisons across massive code repositories. These scalability and cost constraints make LLM-based approaches impractical for industrial deployment in real-world mini-game plagiarism detection scenarios.

\begin{figure}[t]
\centering
\includegraphics[width=0.45\textwidth]{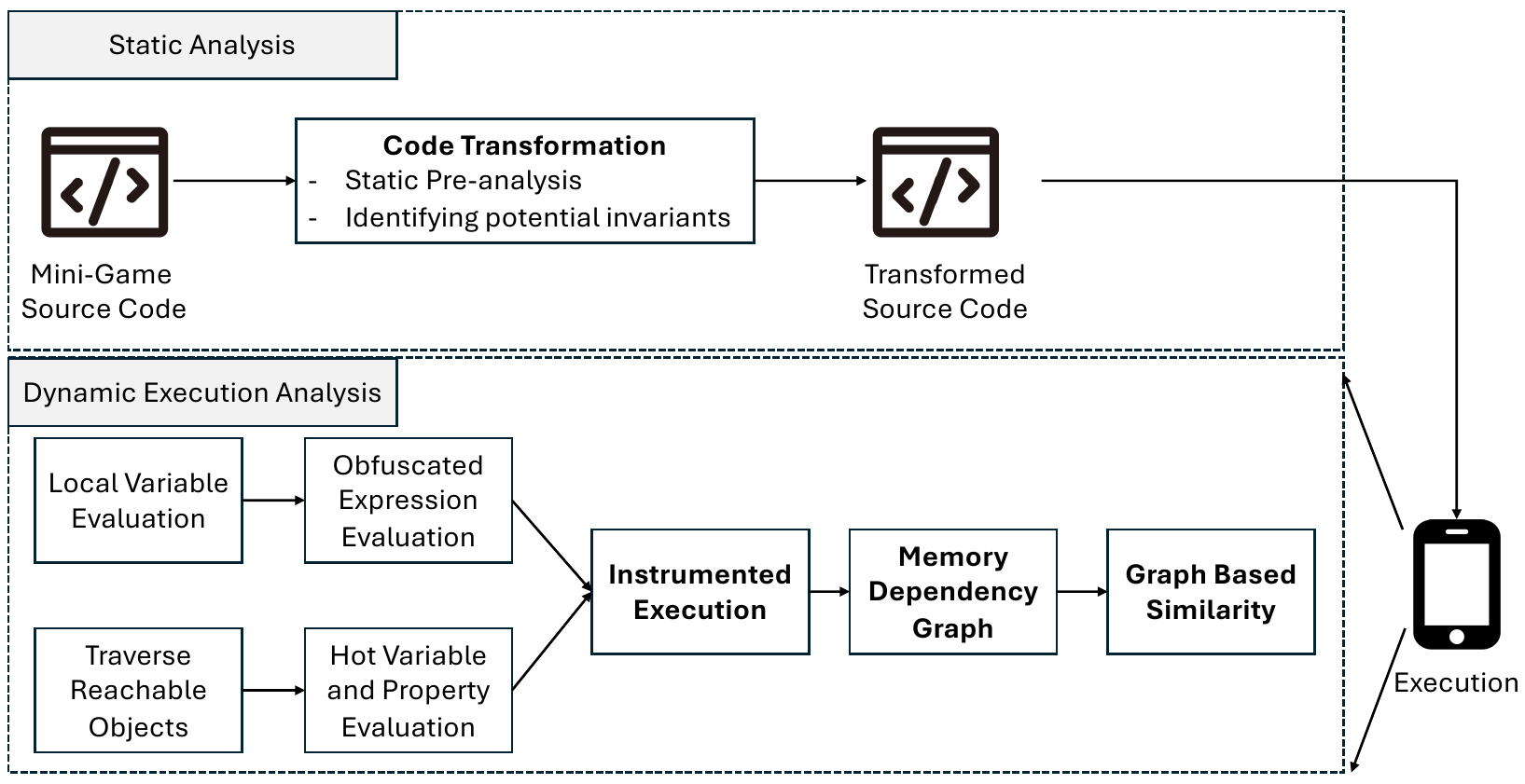} 
\vspace{-4pt}
\caption{The Architectural Overview of JSidentify-V2.
}
\label{fig:framework}
\vspace{-12pt}
\end{figure}
\section{Methodology}

\begin{table*}[h]
\centering
\caption{Comparison with Industry-Known Obfuscation Detection Systems. \CBrush\ indicates full support, \PBrush\ indicates partial support, and \XBrush\ indicates no support.}
\resizebox{\linewidth}{!}{
\begin{tabular}{|c|c|c|c|c|c|c|c|c|}
\hline
\multirow{2}{*}{\textbf{Tool}} & \multirow{2}{*}{\textbf{Copy Paste}} & \multirow{2}{*}{\textbf{Variable Renaming}} & \multirow{2}{*}{\textbf{Dead/Junk Code}} & \multicolumn{3}{c|}{\textbf{Mainstream Obfuscation Tools}} & \multicolumn{2}{c|}{\textbf{Code Decryption}}\\
\cline{5-9}
& & & & \textbf{Control Flow-based} & \textbf{Property-based} & \textbf{String Encryption} & \textbf{Local Key} & \textbf{Cloud Key} \\
\hline
jscpd & \CBrush & \PBrush & \XBrush & \XBrush & \XBrush & \XBrush & \XBrush & \XBrush \\
\hline
JsInspect & \CBrush & \CBrush & \XBrush & \XBrush & \XBrush & \XBrush & \XBrush & \XBrush \\
\hline
GooglePlay & \CBrush & \CBrush & \CBrush & \XBrush & \XBrush & \XBrush & \XBrush & \XBrush \\
\hline
Standford Moss & \CBrush & \CBrush & \XBrush & \XBrush & \XBrush & \XBrush & \XBrush & \XBrush \\
\hline

JSidentify & \CBrush & \CBrush & \CBrush & \PBrush & \PBrush & \PBrush & \XBrush & \XBrush \\
\hline
JSidentify-V2 & \CBrush & \CBrush & \CBrush & \CBrush & \CBrush & \CBrush & \CBrush & \CBrush \\
\hline
\end{tabular}
}
\vspace{-10pt}
\label{tab:compare}
\end{table*}

JSidentify-V2 employs a multi-stage pipeline that systematically transforms the runtime behavior of mini-games into robust, structured fingerprints for plagiarism detection. The fundamental principle underlying our approach is to capture \emph{what} a program accomplishes during execution, rather than relying on \emph{how} its source code is superficially structured. This paradigm shift enables our framework to maintain detection effectiveness even when confronted with sophisticated obfuscation techniques that severely distort static code characteristics. As shown in Table \ref{tab:compare}, JSidentify-V2 supports all kinds of obfuscation methods.

To optimize computational efficiency and resource utilization, our framework adopts a hierarchical detection strategy that prioritizes cost-effectiveness. Initially, we perform static analysis to identify straightforward plagiarism cases, such as instances where code bases are identical or exhibit minimal superficial modifications. Only applications that are deemed non-plagiarized by static analysis alone proceed to the subsequent dynamic analysis phase. This tiered approach ensures that computational resources are allocated judiciously, reserving the more intensive dynamic analysis for cases that genuinely require sophisticated detection capabilities.

Figure~\ref{fig:framework} presents the comprehensive architectural overview of our proposed framework, encompassing four distinct phases that collectively constitute our detection methodology when dynamic analysis is warranted.

\subsection{Formalizing Memory Invariants}

Having established the overall framework architecture, we now detail the theoretical foundation underlying our dynamic analysis approach. The core innovation of our methodology lies in the concept of \textbf{memory invariants}, which serve as robust fingerprints that remain stable even under sophisticated obfuscation techniques commonly employed in mini-game plagiarism.

We define a memory invariant as a runtime artifact whose value or relational property remains consistent across different executions of programs that share the same core logic, despite extensive code obfuscation. This concept is particularly relevant in the mini-game context, where JavaScript code often undergoes aggressive transformations such as variable renaming, control flow flattening, and dynamic code generation through \texttt{eval()} or network-based decryption mechanisms that are prevalent in mini-game platforms.

To formalize this concept, we first identify potential invariant-generating expressions during a static pre-analysis phase. These expressions are typically involved in core computational logic, such as arithmetic operations, function call return values, and key assignments that define the mini-game's core logic (e.g., score calculations, game state transitions, and physics computations).

Let $\mathcal{E}$ be the set of all such expressions in the program's AST. For each expression $e \in \mathcal{E}$, we assign a unique, deterministic location identifier $l_e$, derived from its structural context within the AST (e.g., a hash of its normalized AST subtree and its parent's path).

During program execution, when an instrumented expression $e$ is evaluated, it generates an \textbf{invariant instance}. We formally define an invariant instance $I$ as a tuple:
$$ I = (l_e, v, \tau, t) $$
where:
\begin{itemize}
    \item $l_e$ is the static location identifier of the expression $e$.
    \item $v$ is the concrete value of the expression evaluated at runtime. To handle complex data types, values are serialized into a canonical string format.
    \item $\tau$ is the data type of the value $v$ (e.g., Number, String, Boolean).
    \item $t$ is the execution order index, indicating the sequential position when the evaluation occurred.
\end{itemize}
The set of all invariant instances $\mathcal{I} = \{I_1, I_2, \dots, I_n\}$ collected during a single execution run constitutes the raw data for our analysis. The objective of the subsequent pipeline stages is to structure this raw data into a meaningful, comparison-ready fingerprint.

\subsection{Static Analysis and Basic Dynamic Instrumentation}

Following the established JSidentify framework approach, our system begins with static analysis to extract structural features from mini-games. The static analysis pipeline performs code normalization by simplifying variable names and identifiers, compressing literals, removing whitespaces and comments, and eliminating dead code. We then parse the normalized source code into an AST and extract key structural elements such as function declarations, control flow structures, and variable assignment patterns. It is worth noting that to avoid significant time consumption during the static analysis phase, we have only adopted the basic static analysis capabilities of JSidentify, rather than fully implementing more time-consuming methods like Winnowing Plus and Scene Tree analysis.

While we build upon traditional static analysis techniques like Winnowing Plus and Scene Tree analysis from JSidentify, modern obfuscation techniques can significantly alter code appearance while preserving runtime behavior. This limitation motivates our dynamic invariant extraction approach that captures behavioral patterns resilient to obfuscation transformations.

Our dynamic instrumentation process operates through four key stages to create comprehensive behavioral fingerprints:

\begin{enumerate}
    \item \textbf{Instrumentation Point Identification}: We traverse the AST to identify candidate expressions that are fundamental to program logic and less susceptible to simple refactoring. Our selection prioritizes binary expressions (e.g., arithmetic and comparison operations), call expression return values from user-defined functions, and right-hand side values of assignment expressions.
    
    \item \textbf{Source-to-Source Transformation}: For each identified expression $e$, we inject monitoring code through AST transformation. An expression like \texttt{a + b} becomes \texttt{\_\_log\_invariant\_\_('l\_e', a + b)}, where \texttt{'l\_e'} is a pre-computed location identifier. This transformation preserves the original program logic while enabling runtime monitoring.
    
    \item \textbf{Runtime Invariant Collection}: During program execution, the instrumented code transparently logs invariant instances as tuples $I = (l, v, \tau, t)$, capturing the location identifier, computed value, the potential data type, and execution order. This creates a behavioral fingerprint that reflects the program's computational essence rather than its syntactic appearance.
    
    \item \textbf{Basic Pattern Extraction}: The collected runtime data undergoes initial processing to identify fundamental behavioral patterns and value distributions. However, standard execution often provides incomplete coverage, missing code segments that may contain plagiarized content.
\end{enumerate}

This basic dynamic approach provides a foundation for behavioral analysis, but faces significant challenges when dealing with sophisticated obfuscation techniques that hide critical code segments in rarely executed branches. To address this fundamental limitation, we develop our core contribution: Coverage Enhancement Slicing.

\subsection{Coverage Enhancement via Adaptive Hot Object Slicing}

A major challenge in dynamic analysis is achieving sufficient execution coverage while maintaining acceptable performance overhead. Plagiarists might hide malicious or plagiarized code within deep, convoluted conditional branches or employ sophisticated obfuscation techniques that render traditional instrumentation ineffective. To address this, we introduce \textbf{Adaptive Hot Object Slicing}, designed to maximize coverage of critical code segments while minimizing instrumentation overhead.

Our strategy is based on the insight that certain program entities, which we term \textbf{hot objects}, carry the main program logic and serve as primary sources of behavioral fingerprints. Hot objects encompass two distinct categories: (1) frequently executed functions and critical data structures, class instances, and module objects that are central to program operation (e.g., decryption modules, game state managers, utility libraries), and (2) identifiers with strong semantic correlation to specific game genres, such as ``gun'' in FPS games, which provide domain-specific behavioral signatures. We observe that attackers often intentionally embed key logic within the first category of hot objects for concealment, while the second category reveals game-specific patterns that remain consistent across plagiarized variants, making both types prime targets for our analysis.

Our approach addresses two critical limitations of existing methods: (1) Cloud-based and File Transfer-based attacks that dynamically load obfuscated code rely heavily on hot objects for execution, and (2) even when different code segments appear vastly different after obfuscation, attackers typically apply the same obfuscation algorithms across multiple functions, creating hidden patterns within hot objects.

\begin{algorithm}[t]
\caption{Adaptive Hot Object Slicing}
\label{alg:adaptive-slicing}
\begin{algorithmic}[1]
\STATE \textbf{Input:} Source code $S$, Hot objects set $\mathcal{H}$, Coverage threshold $\theta$, Max rounds $R_{max}$
\STATE \textbf{Output:} Comprehensive invariant set $\mathcal{I}_{comprehensive}$

\STATE // \textit{Stage 1: Initial Hot Object Instrumentation}
\STATE $AST \leftarrow \text{Parse}(S)$
\STATE $\mathcal{O}_{hot} \leftarrow \text{IdentifyHotObjects}(AST, \mathcal{H})$
\STATE $\mathcal{I}_{initial} \leftarrow \emptyset$

\FOR{each $o \in \mathcal{O}_{hot}$}
    \STATE $pos \leftarrow \text{RandomChoice}(\{\text{before}, \text{after}\})$
    \STATE $S \leftarrow \text{InstrumentObject}(S, o, pos)$
\ENDFOR

\STATE // \textit{Stage 2: Initial Execution and Coverage Assessment}
\STATE $\mathcal{I}_{initial} \leftarrow \text{ExecuteInstrumented}(S)$
\STATE $coverage \leftarrow \text{CalculateCoverage}(\mathcal{I}_{initial}, \mathcal{O}_{hot})$

\STATE // \textit{Stage 3: Adaptive Expansion with Round Limit}
\STATE $\mathcal{O}_{uncovered} \leftarrow \{o \in \mathcal{O}_{hot} : \text{GetCoverage}(o) < \theta\}$
\STATE $\mathcal{I}_{expanded} \leftarrow \emptyset$
\STATE $round \leftarrow 0$

\WHILE{$\mathcal{O}_{uncovered} \neq \emptyset$ \textbf{and} $round < R_{max}$}
    \STATE $round \leftarrow round + 1$
    
    \FOR{each $o \in \mathcal{O}_{uncovered}$}
        \STATE $parent \leftarrow \text{GetParentNode}(o, AST)$
        \STATE $S \leftarrow \text{InstrumentParentScope}(S, parent)$
    \ENDFOR
    
    \STATE $\mathcal{I}_{round} \leftarrow \text{ExecuteInstrumented}(S)$
    \STATE $\mathcal{I}_{expanded} \leftarrow \mathcal{I}_{expanded} \cup \mathcal{I}_{round}$
    \STATE $\mathcal{O}_{uncovered} \leftarrow \text{UpdateUncovered}(\mathcal{O}_{uncovered}, \mathcal{I}_{round}, \theta)$
\ENDWHILE

\STATE $\mathcal{I}_{comprehensive} \leftarrow \mathcal{I}_{initial} \cup \mathcal{I}_{expanded}$
\STATE \textbf{return} $\mathcal{I}_{comprehensive}$
\end{algorithmic}
\end{algorithm}

As detailed in Algorithm \ref{alg:adaptive-slicing}, our approach operates in three stages with built-in efficiency controls. First, we identify hot objects within the program and apply lightweight instrumentation at randomly selected positions (before or after object invocation) to minimize predictability for attackers. This initial instrumentation focuses on capturing behavioral patterns from the most critical program components while maintaining low overhead.

Second, we execute the instrumented code and assess coverage for each hot object. Objects that fail to reach a predefined coverage threshold $\theta$ are marked for expanded analysis. This threshold-based approach ensures that we focus additional resources only where needed, maintaining efficiency while ensuring comprehensive coverage.

Third, for hot objects with insufficient coverage, we adaptively expand the instrumentation scope to their parent nodes in the AST, subject to a maximum round limit $R_{max}$ (typically set to 5). This constraint prevents excessive instrumentation overhead while still allowing sufficient exploration of complex obfuscation patterns. The gradual expansion strategy captures hidden patterns that may be distributed across related code segments, particularly effective when attackers apply consistent obfuscation algorithms across multiple functions within the same module or class hierarchy.

This adaptive approach offers several key advantages: (1) it dynamically adjusts instrumentation density based on observed coverage while respecting computational constraints; (2) it can evolve with program versions by redefining hot objects and adjusting instrumentation strategies based on feedback; (3) it specifically targets the execution patterns exploited by Cloud-based and File Transfer-based attacks, where critical logic is dynamically loaded through hot objects; and (4) it captures hidden patterns in obfuscated code by expanding analysis scope when initial coverage is insufficient, while preventing runaway instrumentation through the round limit.

The randomized instrumentation positioning further enhances robustness against adversarial detection, while the hierarchical expansion with bounded iterations ensures that even sophisticated obfuscation techniques cannot evade detection without imposing excessive computational overhead on the analysis process.

\subsection{Memory Dependency Graph (MDG) Construction}

The comprehensive set of invariant instances $\mathcal{I}_{comprehensive}$ collected through our adaptive hot object slicing represents a rich but unstructured collection of behavioral data points. To capture the program's underlying logical structure and leverage the critical insights from hot objects, we organize these instances into an MDG. The MDG provides a powerful, abstract representation of data flow and control dependencies that remains resilient to obfuscation transformations.

We formally define an MDG as a directed graph $G = (V, E, W_V, W_E)$, where:
\begin{itemize}
    \item $V$ is the set of vertices, where each vertex $v_i \in V$ corresponds to a unique instrumentation location $l_i$ from our adaptive slicing process.
    \item $E$ is the set of directed edges representing dependencies between instrumentation points.
    \item $W_V$ assigns feature vectors to vertices, capturing both local properties and hot object membership.
    \item $W_E$ assigns weights and types to edges, representing dependency strength and nature.
\end{itemize}

\textbf{Hot Object-Aware Vertex Construction}: The vertex set $V$ is constructed from unique location identifiers collected during our adaptive instrumentation process. For each unique location $l$ from invariant instances $(l, v, \tau, t) \in \mathcal{I}_{comprehensive}$, we create a vertex. The feature vector $W_V(v)$ incorporates both traditional properties and hot object insights. Syntactic properties include expression type (e.g., \texttt{BINARY\_EXPRESSION}, \texttt{CALL\_EXPRESSION}) and AST-level characteristics. Behavioral statistics encompass the distribution of runtime values $\{v | (l, v, \tau, t) \in \mathcal{I}_{comprehensive}\}$, including mean and variance for numerical invariants, and canonical representations for complex data types. Additionally, we encode hot object membership information indicating whether the location belongs to a hot object and its relative importance within the hot object hierarchy. Coverage metrics record the round number during which this location was successfully instrumented, indicating its accessibility depth within the program structure.

\textbf{Multi-Level Edge Construction}: We establish dependencies through a multi-layered approach that reflects both traditional data flow and hot object relationships:

\begin{enumerate}
    \item \textbf{Intra-Hot Object Dependencies}: Within hot objects, we apply fine-grained data dependency analysis since these components carry the most critical program logic. Direct data dependency edges $(v_i, v_j)$ are created when expression $v_j$ directly uses variables or memory locations modified by expression $v_i$.
    
    \item \textbf{Inter-Hot Object Dependencies}: Between different hot objects, we establish coarser-grained dependencies based on calling relationships and shared data access patterns identified during our adaptive instrumentation.
    
    \item \textbf{Temporal Proximity Dependencies}: For locations instrumented in the same adaptive round, we employ temporal proximity heuristics. An edge $(v_i, v_j)$ is created if instances are logged sequentially within related execution contexts, with edge weights $W_E(e_{ij})$ reflecting both temporal proximity and hot object significance.
    
    \item \textbf{Coverage-Based Dependencies}: Edges connecting locations discovered in different instrumentation rounds capture the hierarchical expansion relationships, providing insights into code organization patterns that persist across obfuscation.
\end{enumerate}

The resulting MDG serves as a hierarchical behavioral fingerprint that prioritizes hot object patterns while maintaining comprehensive coverage. This structure abstracts away from surface-level code characteristics while preserving the essential logical relationships that attackers cannot easily eliminate without fundamentally altering program functionality. The hot object-centric construction ensures that the most critical behavioral patterns receive appropriate emphasis in the final representation.

\subsection{Hot Object-Prioritized Graph Similarity Detection}

With programs represented as MDGs, the plagiarism detection problem is transformed into a graph similarity problem that leverages the hierarchical importance established through our adaptive hot object slicing. Given an MDG $G_S$ from a suspect mini-game and an MDG $G_O$ from an original mini-game in our database, we compute a similarity score $Sim(G_S, G_O)$ that prioritizes critical behavioral patterns while maintaining comprehensive coverage.

A full graph isomorphism or edit distance computation is NP-hard~\cite{bunke1997relation,zager2008graph}. We therefore develop a more efficient, three-stage approach that provides a robust approximation of graph similarity while emphasizing the significance of hot object patterns and multi-level dependencies identified during our adaptive instrumentation process.

\textbf{Stage 1: Hot Object-Weighted Vertex Matching}

We begin by finding optimal matching between vertices of $G_S$ and $G_O$ with explicit consideration of hot object membership. The similarity between two vertices, $v_s \in V_S$ and $v_o \in V_O$, is calculated based on their feature vectors $W_V(v_s)$ and $W_V(v_o)$, incorporating syntactic properties, behavioral statistics, hot object membership, and coverage metrics. The vertex similarity score $S_V(v_s, v_o)$ applies different weighting schemes based on hot object significance. Vertices belonging to hot objects receive higher importance weights, reflecting their critical role in program behavior. We solve this as a maximum weight bipartite matching problem to find the set of matched vertex pairs $M = \{(v_s, v_o)\}$ that maximizes the total weighted similarity. The overall node-level similarity incorporates hot object importance:
$$ Sim_{node}(G_S, G_O) = \frac{\sum_{(v_s, v_o) \in M} w_{hot}(v_s, v_o) \cdot S_V(v_s, v_o)}{\sum_{v_s \in V_S} w_{hot}(v_s) + \sum_{v_o \in V_O} w_{hot}(v_o)} $$
where $w_{hot}(v)$ represents the hot object importance weight of vertex $v$.

\textbf{Stage 2: Multi-Level Structural Consistency}

We evaluate structural consistency across the four types of dependencies established during MDG construction. For each matched pair $(v_s, v_o) \in M$, we compare their local neighborhoods considering intra-hot object dependencies, inter-hot object dependencies, temporal proximity dependencies, and coverage-based dependencies. The neighborhood similarity $S_N(v_s, v_o)$ weighs different edge types according to their significance in preserving program semantics. Intra-hot object edges receive the highest weight as they capture the most critical behavioral patterns, while coverage-based edges provide additional structural validation. 
We define the weighted neighborhood score for a matched pair as:
$$N_{weighted}(v_s, v_o) = \sum_{e \in E_{types}} w_e \cdot S_{N,e}(v_s, v_o)$$
where $E_{types} = \{intra, inter, temporal, coverage\}$. The overall structural similarity is then:
$$Sim_{struct}(G_S, G_O) = \frac{\sum_{(v_s, v_o) \in M} N_{weighted}(v_s, v_o)}{|M| \cdot \sum_{e} w_e}$$

\textbf{Stage 3: Hot Object Pattern Consistency}

We introduce an additional validation stage that specifically examines the consistency of hot object interaction patterns. This stage analyzes whether the overall hot object topology and inter-dependencies are preserved between the suspect and original programs. We compute hot object-level similarity by aggregating the behavioral patterns within each hot object and comparing the resulting signatures. This provides a higher-level validation that complements the fine-grained vertex and edge matching performed in the previous stages.

\textbf{Final Plagiarism Score}

The final similarity score combines all three levels of analysis:
$$ Sim(G_S, G_O) = \alpha \cdot Sim_{node} + (1 - \alpha) \cdot Sim_{struct}  $$
where $\alpha$ is the hyperparameter balancing the importance of individual invariant properties and structural relationships. If $Sim(G_S, G_O)$ exceeds a predetermined threshold $\zeta$, the suspect application is flagged as potential plagiarism. This threshold is determined empirically based on validation sets that include various obfuscation techniques targeting both traditional and hot object-based attack vectors.

\section{Experimental Setup}

\subsection{Evaluation Benchmarks}

For the evaluation of JSIdentify-V2, we construct comprehensive benchmarks across different scenarios.

\textbf{Obfuscation Resistance Benchmark.} We select 50 representative mini-games from WeChat's repository and apply each of the eight obfuscation methods described in Section~\ref{sec:obfuscation} to generate obfuscated versions. This results in 400 obfuscated code samples (50 games × 8 obfuscation methods), which are paired with their original versions to evaluate similarity detection performance under various obfuscation techniques.

\textbf{Real-world Plagiarism Benchmark.} We randomly select 500 pairs of confirmed plagiarism games from WeChat's repository of plagiarism cases, along with 200 non-plagiarism games that have been manually verified not to be involved in plagiarism incidents. We conduct all pairwise combinations among these 1,200 games (i.e., pairing each of the 1,200 games with each of the remaining 1,199 games) to form our comprehensive evaluation dataset.

\subsection{Baselines}

Due to the high evaluation cost, we first collect an initial set of related state-of-the-art approaches and compare their effectiveness on mini-game plagiarism detection, then select the most effective ones as baselines for comprehensive evaluation with JSIdentify-V2.

Our baseline methods include four established JavaScript code similarity detection tools: MOSS, JSCD(safe), Jsinspect, jscpd, and the work of Chen et al. in detecting clones in Android markets \cite{chen2014achieving}. Additionally, we include the previous SOTA method, JSIdentify~\cite{xia2020jsidentify}, to demonstrate the improvements achieved by our enhanced method. For each approach, we set the threshold value that achieves the highest F1-score overall.
\subsection{Evaluation Metrics}
\subsubsection{Identifier \& Property Name Recovery Rate}

This metric evaluates plagiarism detection robustness by measuring the recovery rate of semantically meaningful identifiers from obfuscated JavaScript code. The evaluation process involves: (1) obfuscating source code using tools such as JS-obfuscator, (2) applying different de-obfuscation techniques, and (3) calculating the proportion of successfully recovered variable names and object properties that retain semantic significance. Variables subjected to irreversible renaming are excluded from the calculation, focusing only on identifiers with recoverable semantic traces.

This metric provides insights into how effectively plagiarism detection algorithms can identify code similarities despite intentional obfuscation. Higher recovery rates enable better similarity detection and facilitate manual inspection by preserving semantic information that aids human reviewers in understanding code relationships. 

\subsubsection{Plagiarism Detection Metrics}

We evaluate the effectiveness of our plagiarism detection approach using three standard classification metrics: precision, recall, and F1-score. These metrics assess the accuracy of our method in identifying code plagiarism cases within the benchmark~\cite{powers2020evaluation,sokolova2009systematic}.

Precision measures the proportion of correctly identified plagiarism cases among all cases flagged as plagiarism by our system, calculated as $\text{Precision} = \frac{TP}{TP + FP}$, where $TP$ represents true positives (correctly detected plagiarism) and $FP$ represents false positives (incorrectly flagged as plagiarism). Recall evaluates our system's ability to identify all actual plagiarism cases, computed as $\text{Recall} = \frac{TP}{TP + FN}$, where $FN$ denotes false negatives (missed plagiarism cases). The F1-score provides a balanced measure combining both precision and recall: $\text{F1} = 2 \times \frac{\text{Precision} \times \text{Recall}}{\text{Precision} + \text{Recall}}$. 

\subsection{Research Questions}

To evaluate our proposed JSIdentify-V2 approach, we formulate three research questions:

\textbf{RQ1:  How effective is JSIdentify-V2 in detecting obfuscated code?} We apply the eight obfuscation methods introduced in Section \ref{sec:obfuscation} to mini-game source code and compare our approach with baseline methods in terms of pairwise similarity between original and obfuscated code, as well as variable name recovery rate.

\textbf{RQ2:  How does JSIdentify-V2 perform in real-world plagiarism detection scenarios?} We evaluate detection accuracy using precision, recall, and F1-score metrics on authentic plagiarism cases from production environments.

\textbf{RQ3:  What is the computational efficiency of JSIdentify-V2?} We measure detection time and scalability with increasing source code sizes compared to baseline approaches.

\subsection{Implementation Details}
For fair comparison, all methods are executed on servers equipped with an AMD 9K84 CPU and 1 TB of memory. To minimize randomness in our efficiency measurements, each method is run three times, and we report the average values. For parameter selection, the $\alpha$ parameter in computing the final plagiarism score is set to 0.5, and the predetermined threshold $\zeta$ for the final similarity score is set to 0.6.

In RQ1, we conduct static obfuscation by two obfuscation tools, including JS-obfuscator \cite{jsobfuscator} and UglifyJS \cite{uglify}. For advanced LKD and CKD, we conduct the obfuscation methods by ourselves.
\section{Experiment Results}

\subsection{RQ1: Obfuscation Resistance}

\begin{table*}[th]
    \centering
    \vspace{-5pt}
    \caption{Pair-wise similarity of different plagiarism detection tools under different obfuscation methods.}
    \begin{tabular}{l|cccccc|cc}
    \toprule
    Methods  &  IM & DCI & CFF & NF & SS & SAE & LKD & CKD\\
    \midrule
     MOSS    & 28.3\% & 33.4\% & 7.9\% & 8.2\% & 16.3\% & 3.5\% & 0.0\% & 0.0\%\\
     Simian &2.5\% & 28.4\% & 6.5\% & 4.7\% & 16.5\% & 5.0\% & 0.0\% & 0.0\% \\
     jscpd & 4.2\% & 38.6\% & 14.4\% & 5.5\% & 17.7\% & 2.9\% & 0.0\% & 0.0\%\\
     Jsinspect & 30.7\% & 32.3\% & 15.9\% & 2.6\% & 12.4\% & 6.5\%\ & 0.0\% & 0.0\% \\
     JSCD (safe) & 98.5\%& 60.9\% & 56.3\% & 13.1\% & 33.0\% & 25.4\% & 0.0\% & 0.0\%\\
     AndroidClone & 87.2\% & 45.6\% & 22.6\% & 10.8\% & 15.5\% & 7.6\% & 0.0\% & 0.0\%\\
     \midrule
     JSidentify & 98.1\% & 99.7\% & 77.5\% & 79.6\% & 96.5\% & 93.2\% & 5.4\% & 3.2\%\\
     JSidentify-V2 & \textbf{99.9}\% & \textbf{99.9}\% & \textbf{98.9}\% & \textbf{99.0}\% & \textbf{99.9}\% & \textbf{99.2}\% & \textbf{92.3}\% & \textbf{91.9}\%\\
     \bottomrule
    \end{tabular}
    \vspace{-6pt}
    \label{tab:sim}
\end{table*}

\begin{table}[t]
    \centering
    \vspace{-4pt}
    \caption{Identifier recovery rate of different plagiarism detection tools under different obfuscation methods.}
    \begin{tabular}{l|cc|cc}
    \toprule
    Methods  &   SS & SAE & LKD & CKD\\
    \midrule
     JSidentify & 93.3\% & 91.5\%& 3.2\% & 2.9\%\\
     \midrule
     JSidentify-V2  & \textbf{99.6}\% & \textbf{99.8}\% &  \textbf{99.2}\% & \textbf{99.1}\%\\
     \bottomrule
    \end{tabular}
    \vspace{-6pt}
    \label{tab:recovery}
\end{table}
\subsubsection{Pair-Wise Similarity Measurement} 
\textbf{Static Obfuscation.} We evaluate six traditional static obfuscation methods, including identifier modification, dead code injection, control flow flattening, nested function, string splitting, and string array encoding. The results are shown in Table \ref{tab:sim}. From the table, we can observe that high-intensity variable name modifications (random hexadecimal strings) completely destroy string-level similarity, rendering most existing tools ineffective. While JSinspect is also AST-based, relying solely on structural similarity proves insufficient when string-level features are heavily obfuscated, achieving only approximately 30\% similarity scores. JSCD and AndroidClone demonstrate relatively better robustness against such obfuscation, achieving similarity scores of 87.2\% and 98.5\%, respectively, benefiting from their more sophisticated AST analysis capabilities.

For dead code injection, which affects both AST structure and string content, existing tools show only moderate similarity detection capability since they lack specific considerations for this obfuscation technique. However, for the other four more sophisticated obfuscation methods, all previous approaches demonstrate significantly low recognition success rates, particularly against nested function obfuscation. The previous method, JSIdentify, despite considering various obfuscation techniques, still exhibits performance degradation on control flow flattening and nested function (below 80\% similarity). In contrast, our proposed JSIdentify-V2 leverages dynamic runtime memory characteristics, achieving over 95\% similarity scores across all static obfuscation methods.

\textbf{Dynamic Obfuscation.} These obfuscation techniques utilize local or cloud-based encryption, completely destroying the original code structure and semantics. Consequently, all existing methods, including JSIdentify, fail catastrophically (approximately 0\% similarity). However, our approach captures deep runtime features during code execution, achieving over 90\% accuracy even under these high-intensity obfuscation scenarios.

\subsubsection{Identifier Recovery} Table \ref{tab:recovery} shows the identifier recovery rates under different obfuscation methods. For traditional obfuscation techniques like String Splitting (SS) and String Array Encoding (SAE), the previous JSidentify method achieves over 90\% recovery rates. However, against advanced decryption-based methods (LKD and CKD), static analysis completely fails with almost 0\% recovery, as the encrypted code remains inaccessible without runtime execution. In contrast, JSidentify-V2 consistently achieves over 99\% identifier recovery across all obfuscation methods by capturing runtime behavior after decryption, demonstrating the superiority of dynamic analysis for handling sophisticated obfuscation techniques.

\subsection{RQ2: Real-World Detection Effectiveness}

\begin{table}[t]
    \centering
    \caption{The best F1-score results (across all the threshold values) in real-world plagiarism detection performance of JSidentify-V2 and baselines.}
    \begin{tabular}{c|llll}
    \toprule
     Methods   & Recall & Precision & F1-score & Avg. Time \\
     \midrule
     Jsinspect  & 64.0\% & 54.8\% &59.0\% & 9.8s/pair \\
     MOSS & 52.8\% &60.8\% &56.5\% &1.3s/pair\\
     JSCD (safe) & 60.6\% & 76.6\% &67.7\% &103s/pair\\

     \midrule
     JSidentify & 72.4\% & 99.2\% &83.7\% &24.6s/pair \\
     JSidentify-V2 & \textbf{99.8}\% & \textbf{100}\% &\textbf{99.9}\% & 3.1s/pair \\
     \bottomrule
    \end{tabular}
    
    \label{tab:real}
\end{table}

Table IV demonstrates the performance of JSidentify-V2 in current real-world mini-game plagiarism detection scenarios. We evaluate against three well-established baselines (MOSS, JSinspect, JSCD) along with the previous generation JSidentify method. The results reveal that contemporary mini-game plagiarism employs increasingly sophisticated obfuscation techniques that challenge existing detection approaches. All baseline methods, including the previous JSidentify, achieve recall rates below 75\%, while also producing false positives when analyzing similar but non-plagiarized games. In contrast, JSidentify-V2 achieves 99.8\% recall with 100\% precision, demonstrating substantial improvement in both detection capability and accuracy. Manual inspection of the single missed case revealed it involved partial code plagiarism with \textbf{extensive secondary development}, confirming the effectiveness and reliability of our JSidentify-V2 framework for real-world deployment.

\begin{figure*}[t]
  \centering
  \includegraphics[width=0.85\linewidth]{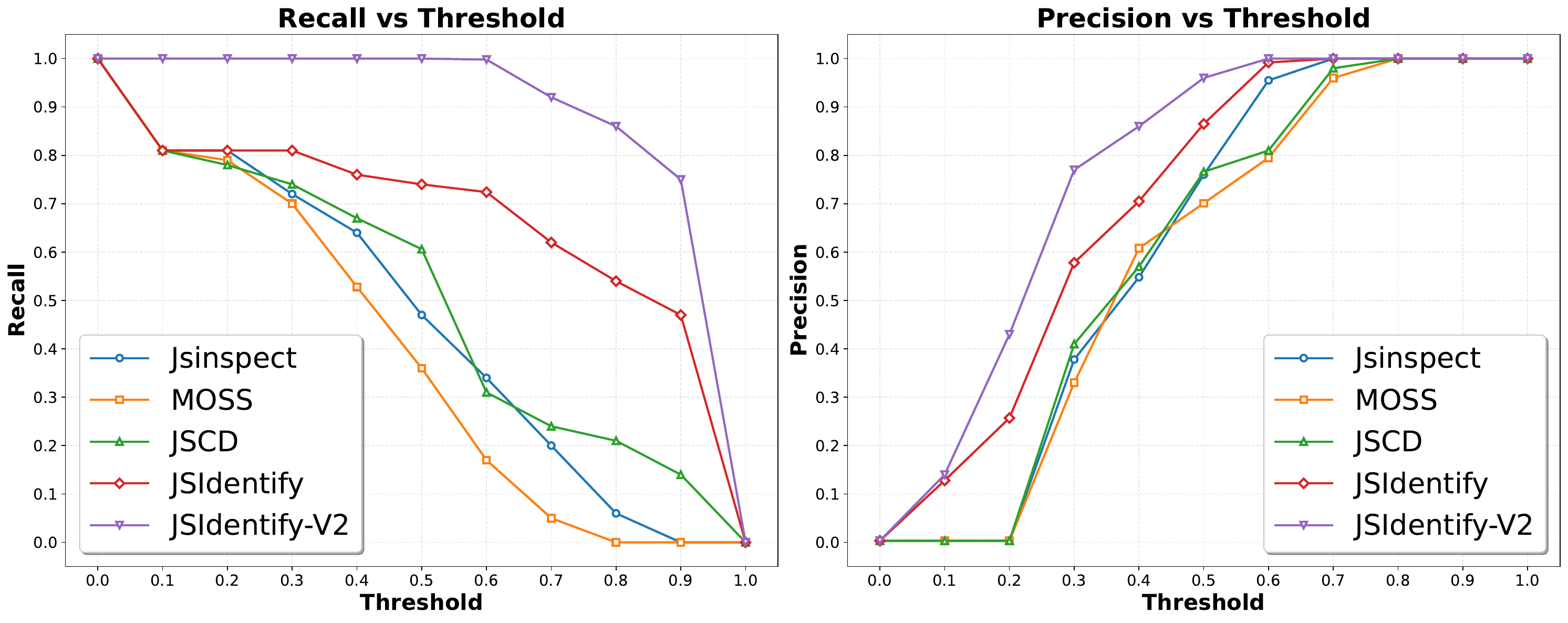}
  \vspace{-4pt}
  \caption{Precision and recall with different threshold values in plagiarism detection.}
  \vspace{-14pt}
  \label{fig:p_r}
\end{figure*}

In addition, we draw Figure \ref{fig:p_r} to illustrate the precision and recall comparison between JSidentify-V2 and baseline methods across different threshold values (ranging from 0 to 1 with a gap at 0.1). JSidentify-V2 demonstrates substantially superior recall performance compared to other approaches. At a threshold of 0.1, while other methods fail to detect deep obfuscation techniques based on Local Key Decryption (LKD) and Cloud Key Decryption (CKD), JSidentify-V2 maintains 100\% recall consistently until the threshold reaches 0.5. In terms of precision, JSidentify-V2 also outperforms other methods significantly. At a threshold of 0.6, our memory fingerprinting-based similarity detection achieves 100\% precision. Overall, JSidentify-V2 delivers superior performance in both recall and precision metrics, resulting in the best F1-score among all evaluated methods and demonstrating its robustness across various threshold configurations.
\subsection{RQ3: Efficiency}
\begin{figure}[t]
  \centering
  \includegraphics[width=0.85\linewidth]{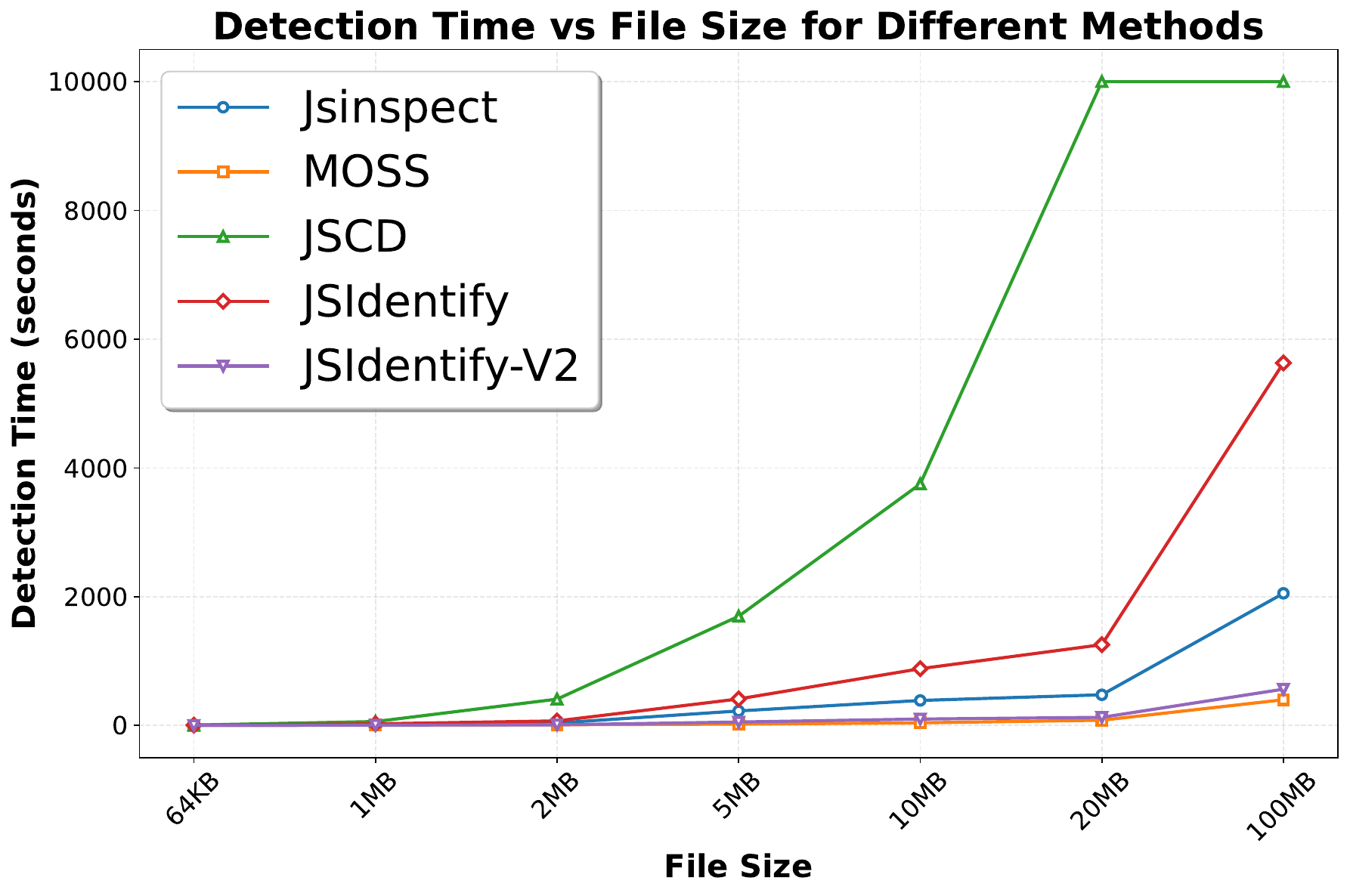}
  \caption{Detection time of JSidentify-V2 and baselines in projects with different sizes of JavaScript source code, where the detection times' reaching 10,000 seconds indicates that the approach is unable to finish the detection within the time limit.}
  \label{fig:time}
  \vspace{-6pt}
\end{figure}
Table \ref{tab:real} also presents the average time consumption for pairwise similarity detection across different methods. JSidentify-V2 demonstrates highly efficient performance with an average detection time of 3.1 seconds per pair, which is approximately 8 times faster than the previous generation JSidentify (24.6s/pair). Compared to other baseline methods, JSidentify-V2 outperforms most approaches in terms of efficiency, being significantly faster than JSCD (103s/pair) and JSinspect (9.8s/pair). While it is slower than MOSS (1.3s/pair), which primarily relies on simple string-based similarity matching, JSidentify-V2 provides substantially better detection accuracy while maintaining competitive runtime performance.

To conduct a more detailed analysis of detection efficiency across different methods, we selected mini-games of varying sizes from WeChat's mini-game repository, ranging from 64KB to approximately 100MB, and compared the detection speeds of JSidentify-V2 against different baselines. As shown in Figure \ref{fig:time}, methods relying entirely on deep static analysis, including JSCD, JSidentify, and JSinspect, exhibit non-linear complexity that approaches quadratic growth with file size. This occurs because static analysis complexity increases dramatically with code volume. Complex mini-games may contain millions of identifiers, making deep static analysis extremely time-consuming.

JSidentify-V2 demonstrates remarkable efficiency through its flexible hierarchical dynamic-static analysis combination. Specifically, when file sizes exceed 20MB, our method requires only one-tenth the time of JSidentify without experiencing a time explosion as file size increases. JSidentify-V2 remains only slightly slower than MOSS, which primarily relies on string similarity, demonstrating the high efficiency of our design approach. While MOSS spends the shortest detection time, it focuses on only a limited number of textual clones.
\section{Discussion}
\subsection{Real-World Deployment Impact}

WeChat currently hosts hundreds of thousands of mini-games with millions of versions in total. With massive numbers of new games and versions uploaded daily, plagiarism detection presents enormous challenges. JSidentify-V2 has been deployed and running stably since June 2024. Even after pre-filtering potential plagiarism cases through similarity hashing, our method is invoked an average of 10 million times daily, comparing uploaded games against the entire repository. To date, we have conducted approximately 4 billion comparisons.

The deployment results have been remarkable. User complaints regarding plagiarism have decreased by over 80\%, dropping from an average of 102 complaints per month to fewer than 20. JSidentify-V2 has become an indispensable tool for protecting the intellectual property rights of WeChat mini-game developers, demonstrating its significant real-world impact in maintaining a healthy development ecosystem.

\subsection{Future Work}
JSidentify-V2 addresses the majority of issues from the previous generation framework~\cite{xia2020jsidentify}, such as resolving source code decryption-based obfuscation methods and flexibly combining static analysis with dynamic execution to improve runtime efficiency significantly. However, we have identified the following challenges as our future work.

First, our current approach may experience reduced recall on mini-games that have undergone extensive secondary development. Even when certain core modules remain unmodified, the overall similarity score tends to be relatively low due to substantial modifications in other parts of the codebase. We plan to develop a modular slice-based plagiarism detection that can precisely identify modular plagiarism within mini-games, enabling more granular detection of copied components while accounting for legitimate modifications.

Second, we aim to actively and continually collect publicly available JavaScript third-party libraries from across the web to build a comprehensive common library database for plagiarism exemption in cases of legitimate code reuse. Many developers legitimately incorporate popular open-source libraries and frameworks into their projects, which should not be flagged as plagiarism. By maintaining an up-to-date database of commonly used libraries and their fingerprints, we can filter out legitimate code reuse and focus detection efforts on actual intellectual property violations.
\section{Conclusion}
In this paper, we have presented JSidentify-V2, a novel dynamic analysis framework that has effectively detected mini-game plagiarism even against sophisticated obfuscation techniques including local and cloud-based decryption methods. By leveraging memory invariants and adaptive hot object slicing, our approach has achieved over 90\% similarity detection across all tested obfuscation techniques while maintaining 99.8\% recall and 100\% precision in real-world scenarios. The framework has demonstrated significant efficiency improvements, delivering an 8$\times$ speedup compared to previous methods, and has proven its practical value through production deployment on WeChat's mini-game platform where plagiarism complaints have decreased by over 80\%. JSidentify-V2 has represented a substantial advancement in protecting intellectual property rights within mini-game ecosystems and has demonstrated the effectiveness of combining dynamic and static analysis for robust plagiarism detection.

\newpage

\bibliographystyle{IEEEtran}
\bibliography{reference,zjnewFull}
\end{document}